\documentclass[amssymb,aps,prl,twocolumn,showpacs]{revtex4}
\usepackage{graphicx}
\usepackage[latin1]{inputenc}
\begin{document}

\title{Motion of an adhesive gel in a swelling gradient: a mechanism for cell locomotion}
\author{Jean-Francois Joanny, $^*$Frank J\"ulicher and Jacques Prost}
\affiliation{Physicochimie Curie (CNRS-UMR168), Institut Curie Section Recherche, 26 rue
d'Ulm, 75248 Paris Cedex 05 France, $^*$Max-Planck Institut f\"ur 
Physik komplexer Systeme, N\"othnitzerstr. 38, 01187 
Dresden, Germany}
\date{\today }

\pacs{87.17Jj, 82.70Gg, 82.35Gh}

\begin{abstract}
Motivated by the motion of nematode sperm cells,  we present a model for the motion of an adhesive gel
on a solid substrate. The gel polymerizes
at the leading edge and depolymerizes at the rear. The motion results from a competition between a self-generated 
swelling gradient and the adhesion on the substrate. 
The resulting stress provokes the rupture of the adhesion points and allows for the motion. The model predicts an unusual force-velocity relation
which depends
in significant ways on the point of application of the force.

\end{abstract}

\maketitle
Many cells are able to crawl on a solid substrate to which they adhere.
This motion  involves in general three processes:
the formation and protrusion of a thin lamellipod in front of the cell,
the adhesion of the lamellipod to the substrate and its retraction at the rear,
pulling the cell forward.
 The formation of the lamellipod involves the 
polymerization of cytoskeletal filaments and their cross-linking 
near the membrane. Motion generation 
requires a symmetry breaking often associated with 
the treadmilling of the cell cytoskeleton (the 
continuous asymmetric polymerization at the leading extremity of the gel and depolymerization at the rear).
Typical examples for crawling cells are the motion of fibroblast cells or fish 
epidermal keratocytes that have been studied in details. 
In many cells the cytoskeleton is an actin
network and the motion generation involves a complex interaction between
actin, myosin II motors and many other proteins 
\cite{mitchison,svitkina1,svitkina2}.

From a 
point of view of physics, cell motion is closely related 
to the motion of an elastic gel. 
Spontaneous motion of gels have been
studied in several contexts. A mechanism has been proposed where 
a gel rolls on a substrate without slipping; the gel is not treadmilling
in this case \cite{degennes1}. 
In order to explain the motion of the bacterium {\em Listeria} which is propelled 
by formation of an actin gel comet, a description has been proposed which is 
based on the interplay of gel elasticity, 
polymerization on the surface of the bacteria
and depolymerization at the outer surface of the gel
\cite{prost}.  
The coupling between polymerization and motion is due to the stress developed between the actin gel and the bacterium. Biomimetic experiments on the motion of small 
 colloidal spheres induced by an actin comet have been performed to test these ideas 
\cite{sykes}. More recently, Bottino et al. \cite{bottino} have described 
the crawling of nematode sperm cells on a surface.
The cytoskeleton is represented by a gel which is treadmilling and adhering to 
the substrate. The polarity 
of the cell leads to a pH gradient that subsequently induces a gradients in
internal tension and adhesion to the substrate which are introduced
phenomenologically.
 
Ascaris nematode sperm cells provide an example of motion of a treadmilling gel in a swelling 
gradient. The motility of these cells has been thoroughly studied experimentally 
\cite{theriot,roberts1}.
The cytoskeleton mostly consists of an apolar  protein, MSP, that similarly to
actin polymerizes to form filaments which can be cross-linked to form  a
gel \cite{king}. The moving cell forms a thin lamellipod on the substrate and the cell body is
dragged behind the lamellipod. There is convincing evidence that 
polymerization of the MSP network proceeds at the advancing edge of a motile cell
\cite{italiano1,roberts2}.
Depolymerization occurs
in the vicinity of the cell body. ATP is hydrolyzed in the 
polymerization process
but molecular motors apparently do not play a role for motility.
The local pH strongly influences the polymerization
and depolymerization of the MSP network\cite{italiano2}: an increase
of pH enhances the polymerization and a decrease of pH provokes 
depolymerization of the network. It has been proposed that an 
influx of protons
in the cell body creates a pH source which leads to a pH gradient and thus an asymmetry
between the front and the rear of the cell\cite{king2}.

The aim of the present letter is to describe on very general physical grounds the motion of 
an isotropic elastic gel adhering on a surface by considering the interplay between 
a self-generated concentration  field (pH)
that creates a gradient in the 
gel properties and the adhesion.  In order to illustrate the main principles of motion of an
adhesive gel in a swelling gradient, we use simplified assumptions. We assume 
that the main effect of the pH is to change 
the local equilibrium swelling of the gel and that
the advancing gel is more contracted at the rear (our mechanism is however more general and does not depend on the sign of the gradient); 
we furthermore assume that the gel adheres to the substrate by localized adhesion points and that the 
adhesion is strongly irreversible. This implies that
the adhesion points are  broken if they are subject to a constraint that exceeds a threshold.
Using general elasticity theory, we show that the combination of adhesion and an 
equilibrium swelling gradient generates stresses in the gel 
that become large at the rear part and provoke  a rupture of the adhesion points.
For the sake of simplicity,  we
use a two-dimensional model, that considers only one dimension in the direction of motion
and the direction  transverse to the plane of motion.  

\begin{figure}
{\centering \resizebox*{8cm}{!}{\includegraphics{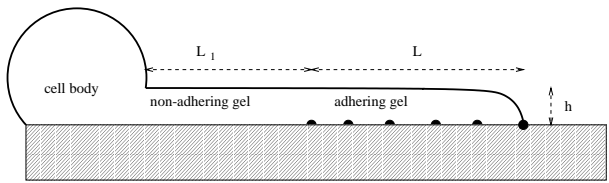}} \par}
\caption{\label{fig: figure1}
Schematic representation of the advancing cell on a substrate.
The lamellipod is filled by a gel that adheres in the front region of length $L$.  
The length of the non-adhering part in 
the back is $L'$. The thickness of the lamellipod is denoted $h$.}
\end{figure}

Consider a thin treadmilling gel in a concentration  gradient (pH). 
The situation sketched in figure (\ref{fig: figure1}) 
involves a gel of small thickness \( h \) that adheres on the 
substrate over a length \( L\gg h \).
Behind the adhering part, there is a non adhering part of the gel of length
\( L' \).
The pH gradient is stationary in the reference frame of the moving gel. 
The gel forms at the leading edge by polymerization and it depolymerizes at the rear. 
In the following, we use the state of the gel formed by polymerization
as the reference state for the quantification of elastic deformations: in this 
state, two points
connected by a vector \( {\bf dx} \) are at a distance \( 
ds_{0}^{2}=\delta _{ij}dx_{i}dx_{j} \).
As the polymerization proceeds, each point of the gel is  
transported  by the treadmilling process inside the moving gel and displaced 
by a vector \( {\bf u} \) with respect to the substrate. The gel
undergoes elastic deformations resulting in changes in the distance between the 
two  points which becomes
\( ds^{2}=g_{ij}dx_{i}dx_{j} \) where \( g_{ij}=\delta 
_{ij}+2\epsilon _{ij} \) is the metric tensor and $\epsilon _{ij}=\frac{1}{2}(\frac{\partial 
u_{i}}{\partial x_{j}}+\frac{\partial u_{j}}{\partial x_{i}})$ is the displacement gradient
(which should not be confused with the strain tensor defined below)
\cite{lubensky}.
 In the reference frame of the gel, a given volume element is transported 
toward low pH regions as time goes on;
its equilibrium swelling changes and it tends to contract. We assume 
here for simplicity
that the equilibrium state of the gel at a given position (a given
pH) is isotropic. Swelling of the gel implies that
the distance between two neighboring points in the equilibrium state
is contracted by a factor \( \Lambda  \) that depends only on the local pH, as
compared to the reference state in which it was formed. 
The contraction factor  \( \Lambda  \)  increases from
a value $\Lambda<1$ at the rear of the gel to $\Lambda=1$ 
at the leading edge. The equilibrium state
of the gel is the state that a small gel element would reach if it were cut
from the rest of the gel while remaining in its environment. The metric tensor in
the local equilibrium state is \( g^{0}_{ij}=\Lambda ^{2}\delta _{ij} \). The
strain tensor is defined in a standard way as \( 
u_{ij}=\frac{1}{2}(g_{ij}-g^{0}_{ij})=\frac{_{1-\Lambda 
^{2}}}{2}\delta _{ij}+\epsilon _{ij} \).
The stress tensor in the gel \( \sigma _{ij} \) can be calculated by assuming
again that the gel is isotropic and by introducing the two Lamé coefficients
\( \lambda  \) and \( \mu  \) such that \( \sigma _{ij}=\lambda 
u_{kk}\delta _{ij}+2\mu u_{ij} \). The two Lamé coefficients may 
themselves
depend on the local pH but, within a linear elasticity description, we ignore here this variation. 
Our model for the motion is one-dimensional, \( x \) is the coordinate parallel to the substrate and \( z \) 
is perpendicular. The stress tensor can then be written as $
\sigma _{ij}=(\lambda +\mu )(1-\Lambda ^{2})\delta _{ij}+
\lambda \epsilon _{kk}\delta _{ij}+2\mu \epsilon _{ij}$.
The second and third terms in this expression are the standard contributions
from linear elasticity theory; the first term, related to the 
change in equilibrium swelling, plays the role of the tensile 
stress introduced
phenomenologically by Bottino et al. \cite{bottino}.

Locally the gel is at mechanical equilibrium and the divergence of the stress
tensor vanishes \( \frac{\partial }{\partial x_{j}}\sigma _{ij}=0 \). However,
the gel does not relax to its equilibrium swelling since it adheres 
on the surface.
If we assume that the adhesion occurs at the leading edge upon the formation
of the gel and that the adhesion points are frozen, the displacement  \({ \bf u }\) vanishes 
at each point on the
surface. The upper surface of the gel is free
and the two components of the stress vanish there \( \sigma _{iz}(h)=0 \).

The elastic deformation is calculated by solving the force balance equation. Inside
the gel, we use a lubrication approximation  
\cite{batchelor}:
the variation of the displacement occurs over a distance $h$ in the
perpendicular direction (\( z\)) much smaller than the distance $L$ in the direction parallel to the substrate
and the derivatives with respect to \( z \) are thus much larger than 
the derivatives
with respect to \( x \). We find \cite{dewimille}
\begin{eqnarray}
 & u_{x}=\frac{\lambda +\mu }{\lambda +2\mu }\frac{d(\Lambda ^{2}(x)-1)}{dx}(\frac{z^{2}}{2}-2hz)& \label{displacement} \\
 & u_{z}=\frac{\lambda +\mu }{\lambda +2\mu }(\Lambda ^{2}(x)-1)z & \nonumber 
\end{eqnarray}

\noindent The deformation in the \( x \) direction at the free surface ($z=h$) is negative towards the
rear where the gel contraction is stronger. The ``lubrication 
approximation''
is valid everywhere in the gel except in a region of size \( h \) both
at the leading edge and at the rear of the gel. The distance to the edge of the gel is smaller than $h$ in these regions and the full elasticity
equations must be solved. At the rear, the gel tends to contract and  the deformation
in the \( x \) direction 
becomes positive. 
The tangential stress on the
substrate ($z=0$) at position \( x \) reads
\begin{equation}
\label{stress}
\sigma _{xz}=-2\mu \frac{\lambda +\mu }{\lambda +2\mu 
}h\frac{d(\Lambda ^{2}-1)}{dx}
\end{equation}

The stress on the surface is negative as the gel is pulled towards the rear
part. The tangential stress on the surface is balanced by the force per unit area exerted
by the adhesion points. If there is a uniform density \( \rho _{s} \) 
of adhesion
points, the (positive) force exerted by each point on the gel  is \( f=-\sigma 
_{xz}/\rho _{s} \).

So far, we have considered a flat piece of gel of length \( L \)
adhering on the substrate. At the rear of this piece over a region with a size
of the order of the thickness \( h \), the signs of both the deformation and
the tangential stress change. A precise calculation of the tangential stress
in this region is difficult as it requires a complete solution of the 
elasticity
equations. However, one can get an estimate of the stress using the macroscopic
force balance on the adhering gel along the substrate. 
The total force acting on the gel vanishes. 
Since there are no external forces acting at the upper surface and at the leading 
edge, the force balance leads to
\( \int _{0}^{L}\sigma _{xz}dx=0 \). If we neglect the variation of 
the tangential
stress in the region of size \( h \) where it is positive, we obtain
\begin{equation}
\label{backstress}
\sigma _{xz}(x=0)\sim \mu \frac{\lambda +\mu }{\lambda +2\mu }(1-\Lambda ^{2})
\end{equation}
The ``lubrication'' approximation is consistent if \( 
h\frac{d(\Lambda ^{2}-1)}{dx} \ll 1 \)
and the stress pulling the gel forward at the back of the gel is much larger
than the stress pulling it backwards in the front parts. The tension on the
adhesion points is thus largest at the back of the adhering part of the gel.

The adhesion points on the substrate have a finite strength,
they rupture if the force exerted by the gel is too large. We consider here that adhesion is irreversible and 
that there is a threshold for the rupture of an adhesion point 
(that can be associated to several molecular bonds). The rupture of each
adhesion point can be viewed as the escape of a particle from an attractive
potential well under the action of an external force (the gel stress). This
problem has been studied in details by several authors \cite{bell,evans}; we 
use here the simplified approach that ignores thermal fluctuations which 
amounts to characterize each adhesion point by the critical force 
\( f_{c} \) that it can sustain (This force could also
in principle depend on pH and on the time scale of the
adhesion breakage \cite{evans}). The stress exerted on the substrate by the
gel being largest at the back of the adhering gel, the adhesion points first
break at the back of the gel. The critical adhesion force, using
equation (\ref{backstress}), 
defines 
a critical value of the tangential stress
and thus a critical value of the swelling ratio \( \Lambda _{c} \) 
at which adhesion bonds rupture. 
If we assume steady state motion, 
the pH profile and 
thus also the profile of swelling \( \Lambda (x) \) are stationary 
in the reference frame of the gel. 
The critical adhesion force
then imposes the distance \( L' \) between the rear of the gel and the point where
the adhesion points rupture when 
\( \Lambda (0)=\Lambda _{c} \). An important point is that our model does not require
any adhesion gradient, the adhesion rupture occurs at the point of strongest tension.

We therefore obtain the picture sketched in fig. (\ref{fig: figure1}) of
a gel adhering to the substrate in the front part over a length \( L \)
but not adhering in the rear over a length \( L' \). The adhering gel is under stress while 
the non-adhering
gel can relax to its local equilibrium swelling characterized by \( 
\Lambda (x) \). 
The picture is not changed qualitatively if the critical force of the adhesion
points \( f_{c} \) depends on pH, only the precise value of \( L' \) changes.
The concept of rupture of discrete adhesion points implies that
the motion is not steady but occurs in jumps. 
As the depolymerization (and the polymerization) proceeds, 
the force on the most strongly loaded adhesion point increases further until 
the critical value is reached which  provokes rupture.  
When an adhesion point ruptures, the gel relaxes
rapidly and moves in the direction of motion. 
We implicitly assume here  that the rupture is  irreversible: 
adhesion points do not reform immediately and the kinetics of formation
of the adhesion points is slow compared to the depolymerization kinetics. 

The model is consistent if the internal stress is large enough to induce
rupture of the adhesion points ($L' $ must be positive) and if the adhesion points can form
($L$ must be larger than $h$) otherwise 
the gel gets stuck on the substrate. Our simplified description relies on the
use a two-dimensional model. For a real 
three-dimensional gel, collective effects such as detachment waves similar to 
those observed in friction problems could be expected \cite{ronsin}. 
This will be the subject of future studies.

The gel adhering on a surface advances by polymerization at
the leading edge and depolymerization at the rear. 
The motion is possible because of the rupture of the adhesion points that
we just discussed. 
The polymerization reactions are far from equilibrium
and we assume for simplicity that only polymerization occurs at the leading edge and only
depolymerization occurs at the rear. The free energy gain in these reactions
(as well as the energy required to maintain the pH gradient) are the 
sources of energy for the motion. The dissipation is mainly due to the gel 
retraction after the rupture of adhesion points.

In order to obtain a complete model for motion, we have to characterize
the polymerization and depolymerization reactions.
We denote  the local concentration of free monomers by \( \rho (x) \), the local gel concentration by
\( c(x)=c_{0}\, \Lambda (x)^{-2} \), where
\( c_{0} \) is the concentration at the leading edge
of the gel. 
The total number of monomers in the gel,  \( m_{gel} \), increases by polymerization
and decreases by depolymerization. 
The kinetics of these reactions is written as
\begin{eqnarray*}
 & \frac{dm_{gel}}{dt}\mid _{l}=h_l k_{p}c_{0}\rho_{l} & \\
 & \frac{dm_{gel}}{dt}\mid _{r}=-h_r k_{d}c_{r}=-k_{d}hc_{0}/\Lambda _{r} & 
\end{eqnarray*}
\noindent Here, the subscripts \( l \) and  \( r \) refer, respectively, to  
the leading edge of the gel where polymerization occurs and the 
the rear where the depolymerization occurs.
The rate constants for polymerization \( k_{p} \)  and depolymerization
\( k_{d} \) should be taken at the local pH (at the leading edge for 
polymerization and at the rear for depolymerization). In a steady 
state, the gel mass is constant and 
$ \frac{dm_{gel}}{dt}\mid _{l} +\frac{dm_{gel}}{dt}\mid _{r}=0 $. 
In the reference
frame of the moving gel, the transport of the depolymerized monomers occurs
by diffusion. In a steady state, the flux \( J=D_{m}(\rho_{r}-\rho_{l})/(L+L') \)
is equal to \( -\frac{dm_{gel}}{dt}\mid _{r}/h \),
(\( D_{m} \) being the monomer diffusion constant). This fixes the monomer concentrations at 
the leading edge and at the rear \(\rho_{l}=k_{d}/(k_{p}\Lambda _{r}) \) and \( 
\rho_{r}=\rho_{l}\left\{ 1+k_{p}c_{0}(L+L')/D_{m}\right\} 
\).
The average velocity is equal to the polymerization speed 
$Vc_0 =\frac1{h} \frac{dm_{gel}}{dt}$ or
\begin{equation}
\label{vitesse}
V=k_{p}\rho_{l}=k_{d}/\Lambda_{r}
\end{equation}
It is selected by the steady state polymerization or depolymerization velocity.
 The total length of the gel \( L+L' \),
can be determined assuming the total monomer mass \( m \) (free and in the gel)
is known. 

The main result of this letter is to show that a gel
can sustain a  steady state motion via a treadmilling mechanism 
even though it adheres to the substrate. No gradient in adhesive strength is  required for this to happen.
The two necessary ingredients are the existence of a critical rupture force for the adhesive bonds 
and of a self-sustained gradient leading to a change in the gel swelling.
The critical rupture force depends on the loading rate, the actual value must therefore 
be determined taking the gel speed into account. 
Furthermore, the existence of the mechanism does not depend on the sign of the induced gel swelling. 
The gradient that controls swelling has been described in terms of a pH gradient,
but any other gradient in the gel could play this role. 
The differential swelling of the gel could also result from the action of molecular motors 
as will be discussed in a forthcoming publication. 
We have presented a 
generic mechanism. Its originality and robustness stem from the peculiar distribution of the stresses 
at the gel-substrate interface. 
Over most of the gel surface, the stress pulling on the adhesive 
bonds is proportional to the gel thickness and 
the swelling gradient. In a small region where the rupture occurs, conventional elasticity tells us 
that it is of opposite sign and larger 
by a factor $L/h$.  For nematodes, $L/h$ is of order $10-20$ and there is a broad range of adhesive strength over which the mechanism can work. This stress distribution 
can be measured using techniques recently developed to study the forces that cells develop on soft substrates\cite{schwarz}. 
A sign inversion of the stress as proposed here 
has been observed on keratocyte cells \cite{sheetz}.

It is very instructive to study within our framework the situation where external forces act on the moving gel.
The effect of a force has a striking signature depending on the position where the force
is applied.  Consider an external force  applied at position $x$ along the substrate. If the force is 
acting at a point inside the adhering gel, the elasticity equations show 
that its effect is screened: the gel deformation and 
the stress on the surface are changed only over a region of
size $h$ (the gel thickness) around the point of application; the force therefore 
has no effect on the gel velocity, total length 
and adhering length until it reaches a bond rupture value. 
If the force is applied at the front of the gel, its effect is also localized. The force however tends to reduce the polymerization rate; the behavior
of the gel in a steady state strongly depends on the depolymerization conditions at the rear of the gel. 
If  the force is applied at the rupture point, it must be added to the macroscopic force balance in the 
gel which determines the stress
at the rupture point. In the presence of an additional force,
the adhering section of the gel is increased 
and corresponds to an effective rupture force  $f_c+f/(\rho_c h)$. 
Eventually, a force applied 
at the rear of the gel and opposing the motion influences the depolymerization. In general it increases the depolymerization rate and according to equation \ref{vitesse}
the velocity increases. The gel in this case has a negative mobility! All these surprising predictions could be tested experimentally.
We will present a more detailed study of the effect of a force in a future publication.

We are grateful to J.Plastino (Institut Curie) for useful 
discussions on the 
MSP protein and A.Mogilner for interesting discussions on ref. \cite{bottino}.

\end{document}